\documentclass[12pt]{article}

\usepackage[english]{babel}
\usepackage{graphicx}
\usepackage{amsfonts}
\usepackage{amsmath}
\usepackage{amssymb}

\usepackage[title,titletoc]{appendix}
\usepackage[footnotesize,bf]{caption}
\usepackage{authblk}
\usepackage{setspace}
\usepackage[left=3cm,top=3cm,bottom=3cm,right=3cm,nohead,nofoot]{geometry}

\begin{document}

\title{Application of a stochastic modeling to assess the evolution of tuberculous and non-tuberculous mycobacterial
infection in patients treated with tumor necrosis factor inhibitors}


\author[1,2]{Elena Agliari\thanks{\mbox{\emph{E-mail~address}:~elena.agliari@fis.unipr.it}}}
\author[3,4]{Lorenzo Asti}
\author[4]{Adriano Barra}
\author[5]{Rossana Scrivo}
\author[5]{Guido Valesini}
\author[6]{Robert S. Wallis}
\affil[1]{\footnotesize Dipartimento di Fisica, Universit\`{a} di Parma, viale G. Usberti 7/A, 43100 Parma, Italy.
\affil[2]INFN, Gruppo Collegato di Parma, viale G. Usberti 7/A, 43100 Parma, Italy.}
\affil[3]{Dipartimento di Scienze di Base e Applicate per l'Ingegneria, Sapienza Universit\`{a} di Roma, Via Antonio Scarpa 16, 00161 Roma, Italy.}
\affil[4]{Dipartimento di Fisica, Sapienza Universit\`{a} di Roma, Piazzale A. Moro 2, 00185 Roma, Italy.}
\affil[5]{Dipartimento di Medicina Interna e Specialit\`{a} mediche, Sapienza Universit\`{a} di Roma, Viale del Policlinico 155, 00161 Roma, Italy.}
\affil[6]{Pfizer,  445 Eastern Point Road,  Groton CT 06340, Connecticut, United States of America}


\date{}

\maketitle

\begin{abstract}
In this manuscript we apply stochastic modeling to investigate the risk of reactivation of latent mycobacterial infections in patients undergoing treatment with tumor necrosis factor inhibitors. First, we review the perspective proposed by one of the authors in a previous work and which consists in predicting
the occurrence of reactivation of latent tuberculosis infection or newly acquired tuberculosis during treatment; this is based on variational procedures on a simple set of parameters (e.g. rate of reactivation of a latent infection). Then, we develop a full analytical study of this approach through a Markov chain analysis and we find an exact solution for the temporal evolution of the number of cases of tuberculosis infection (re)activation.
The analytical solution is compared with Monte Carlo simulations and with experimental data, showing overall excellent agreement.
The generality of this theoretical framework allows to investigate also the case of
non-tuberculous mycobacteria infections; in particular, we show that reactivation in that context plays a minor role.
This may suggest that, while the screening for tuberculous is necessary prior to initiating biologics, when considering non-tuberculous mycobacteria only a watchful monitoring during the treatment is recommended.
The framework outlined in this paper is quite general and could be extremely promising in further researches on drug-related adverse events.
\end{abstract}



\newcommand{\beq}{\begin{equation}}
\newcommand{\eeq}{\end{equation}}
\newcommand{\bea}{\begin{eqnarray}}
\newcommand{\eea}{\end{eqnarray}}
\newcommand{\np}{n_{\bf p}({\bf r},\,t)}
\newcommand{\vp}{{\bf v}_{\bf p}}
\newcommand{\ep}{\epsilon_{\bf p}}
\newcommand{\primo}{^\prime}
\newcommand{\sigsig}{({\bm \sigma}_1 \cdot {\bm \sigma}_2)}
\newcommand{\tautau}{({\bm \tau}_1 \cdot {\bm \tau}_2)}
\def\openone{\leavevmode\hbox{\small1\kern-3.8pt\normalsize1}}
\def\<{\langle}
\def\>{\rangle}
\def\lsim{\buildrel < \over {_{\sim}}}
\def\gsim{\buildrel > \over {_{\sim}}}
\def\dslash{\partial\!\!\!/}
\def\slash{\!\!\!/}
\def\sla{\!\!\!\!/}
\def\E{\mathbb{E}}
\def\ap{\left(\frac{1+a}{2}\right)}
\def\am{\left(\frac{1-a}{2}\right)}
\def\sumc{\sum_{\{\sigma\}} \int \prod_{\mu} d\mu(z_\mu)}
\def\sumimu{\sum_{i,\mu}^{l_{\eta},l_{\chi}}}
\def\sumi{\sum_{i}^{l_{\eta}}}
\def\summu{\sum_{\mu}^{l_{\chi}}}
\def \ap{\bigg(\frac{1+a}{2}\bigg)}
\def \g{\frac{\gamma}{2}}
\def \sg{\sqrt{\frac{\gamma}{2}}}

\section{Introduction}

Over the last decades the improved understanding of the pathogenesis of chronic inflammatory diseases, together with a major advance in biotechnology, have accelerated the development of biological therapies, designed to neutralize specific targets that mediate and sustain the clinical manifestations of diseases. These compounds, mainly monoclonal antibodies (mAb) and fusion proteins, introduced a breakthrough in the management of different conditions including inflammatory rheumatologic disorders \cite{tracey}.
In this context, the first setting of application of the biological agents was rheumatoid arthritis (RA), a chronic autoimmune disease affecting approximately $1\%$ of the adult population \cite{lee}. If the disease is not treated adequately, progressive deformity can lead to loss of quality of life and reduce average life expectancy by about a decade \cite{lee}. Studies on the pathogenic mechanisms of RA have revealed that tumor necrosis factor (TNF) is a cytokine playing a critical role in the inflammatory cascade that results in the irreversible joint damage typical of the disease \cite{choy}. Following these discoveries, a series of clinical trials in patients with RA showed the therapeutic benefit of TNF blockade \cite{elliott}. As a consequence, five biological agents engineered to block TNF actions are currently available: infliximab, adalimumab, golimumab, certolizumab pegol (all of them mAb), and etanercept (a receptor fusion protein) \cite{tak}. While being highly effective, TNF blockers have raised concerns about the potential for an increased susceptibility to infections, in particular the reactivation of latent tuberculosis (TB) infection \cite{lancet,Wallis_bio,Wallis_lancet,Wallis_math,Wallis_data}.
Mycobacterium tuberculosis, the cause of human TB, can result in a metastable clinical latency lasting for decades. 
Much has been speculated about the structure of granuloma which should contain Mycobacteria, since murine models indicated that TNF was necessary for both formation and maintenance of granulomas \cite{plessner}. However, subsequent studies on zebrafish model \cite{clay}, monkeys \cite{lin}, and humans \cite{keane,iliopoulos,liberopoulos,verhave} challenged these data, demonstrating that the crucial role of TNF in the granuloma was indeed macrophage activation and stimulation of chemokine production.
The reactivation of latent TB infection has been associated with all TNF inhibitors, hence pre-initiation screening procedures have been recommended, which have successfully reduced the number of reported cases \cite{schiff}, although current screening tools lack sensitivity and specificity \cite{Scrivo-ClinRheum2012,Scrivo2}.

TNF blockers seem to increase also the risk of other granulomatous diseases, but little is known about the emergence of illness due to non-tuberculous Mycobacteria (NTM).
These are a huge ensemble of pathogens (e.g. M. avium, M. abscessus, and so on \cite{Wagner})  and up to date, approximately $50$ different mycobacterial species are considered to be etiological agents of human
diseases and this number seems  still growing \cite{Wagner}.
Most cases these days occur in hosts with relatively intact immune responses. 
However, RA and other chronic diseases with pulmonary manifestations can predispose a person to NTM pulmonary disease \cite{griffith} expressing as a possible serious complication, especially in immunosuppressed subjects. Thus, it is of utmost importance to study also the risk related to NTM, in the perspective to understand if a proper screening may be helpful in conferring a wider protection to the patients. This is particularly true, in that the TNF blockers appear to predispose both to disseminated and localized disease \cite{Winthrop,Wagner}, but also because these infections are increasing in prevalence, especially among women, which are more frequently affected by RA than men.
%
In the present state of the art, the scenario for NTM diseases,
with respect to TNF-blocking drugs, seems different from the TB counterpart: in particular, through extensive experimental screening, both Wallis and coworkers \cite{Wallis_data}, and  Winthrop and coworkers \cite{Winthrop,Winthrop2}  evidenced that latency and reactivation do not seem to  play a crucial role in this context, yet a clear-cut picture is still missing.

Now, as far as TB is concerned, data collected through the Adverse Event Reporting System of the US Food and Drug Administration (FDA)
in the time-window $1998-2002$, related to the two
test-case drugs with a different mechanism of action, i.e. infliximab and etanercept, highlight that
TB infections involve $54$ over $10^5$ patients treated with infliximab and $28$ over $10^5$ patients
receiving etanercept \cite{Wallis_data}. Therefore, the question is: As the latency in TB can last decades,
are these infections (in patients under therapy) new ones or are they reactivation of previously encountered pathogens due to a suppressed immune system?
This kind of question underlies the awareness of a real need and disposal for extensive pre-screening procedures.
Unfortunately, the answer is by far not trivial as, for TB, there are no secure pathways to discriminate between a new infection
or the raise of a previous one. Moreover, a clear methodology for finding latencies is still lacking.
Furthermore, the rarity and different sizes of this infection in different countries (ranging from $5$ over $10^5$ in Sweden up to $140$ over $10^5$ in Romania \cite{Wallis_lancet}) implies that data analysis and its
subsequent interpretation must be carefully performed.

As for NTM, still from FDA, through the post-marketing surveillance system (MedWatch)
and through a further survey within the Emerging Infections Network of the Infectious Diseases Society of
America (IDSA), Winthrop and coworkers reported a detailed study of possible correlations between
the usage of TNF blockers and the emergence of NTM diseases: over a time-window of $8$ years, they highlighted
a higher prevalence of NTM diseases in patients treated with infliximab rather than etanercept \cite{Winthrop2}.

In order to investigate possible correlations between the incidence of infections by such Mycobacteria  and biological therapy, one could rely directly on the molecular details of TNF processing signal (which has been, at least partially, elucidated, see e.g. \cite{moleculardetails}), coupled to the underlying infliximab and etanercept mechanism of action, which could be achievable directly through molecular immunology approach. Beyond these ``standard" strands, a completely different route can also be performed:
Given the relative large amount of collected data, the problem can be considered from a purely inferential viewpoint, by-passing the underlying molecular immunology know-how (see also \cite{marino,joanne}). According to this perspective, in Ref. \cite{Wallis_math, Wallis_data}, an abstract (logical) environment for TB case has been defined, where patients can occupy
one of the (following) five different states: $(0)$ No infection, $(1)$ New infection, $(2)$ Latency, $(3)$ Reactivation of a previous TB infection, $(4)$ Post first TB encounter.
Clearly, the patients starting the therapy (and hence belonging to the survey)
can correspond to either state $(0)$ or $(2)$, because all the other states imply quantifiable sickness and the patient would then be treated for TB rather than RA. Then, at the end  of the survey, a fraction of these patients will be in an illness state, i.e. either state $(3)$ or $(4)$.
The transition rates between different states are assumed as free-parameters, whose values are estimated through numerical simulations: the best estimate is the one able to reproduce, with the smallest error, the experimental data. Remarkably, the probability of latent TB reactivation in patients treated with infliximab turned out to be an order of magnitude per unit of time higher than the same probability for patients trated with etanercept \cite{Wallis_math}.

Here, we first formalize this approach in terms of Markov chains and we write the related Master equation in continuous-time limit, then we solve the model analytically and study its properties in full details. In this way we get the explicit expression for the number of patients $c(t)$ exhibiting a TB (re)activation, as a function of time $t$.
One step forward, we check the robustness of our results through extensive Monte Carlo simulations and over the clinical data of the TB scenario, finding overall excellent agreement among all our results (and previous literature).
Moreover, we find that different magnitudes for the probability of reactivation correspond to qualitative different behaviors for $c(t)$ (on the proper timescale), that is, the number of patients displaying active infection increases exponentially in time when using infliximab and linearly in time when using etanercept.

The analytical expression for the whole evolution of the system implies a great feasibility of the technique itself (e.g. we have the whole set of first integrals and a clear picture of all the hidden symmetries) and also allows to address, in complete generality, several instances. In particular, we can finally consider generic NTM infections, where, interestingly, the scenario appears quite different from the TB counterpart: clinical data suggest that $c(t)$ (on the proper timescale) grows quadratically with time and this is recovered by our analytical picture only under the assumption of a negligible role played by latency reactivation. We check these findings also through extensive Monte Carlo runs, which are in full agreement too. Remarkably, this is very consistent with the present state of the art in the medical literature dealing with NTM.

As a final result, there are two types of conclusions which stem from our work: The former belongs to the world of modelers, while  the latter to the world of clinicians.
\newline
From a mathematical perspective, the encouraging results of this approach may pave the way for the development of handily and fruitful instruments for physicians.
\newline
Much more carefully, in the clinician's counterpart, as this approach bypasses the whole underlying biological complexity, it may contribute to confirm, from a theoretical perspective, the current understanding of adverse events coupled to TNF-inhibitors and the consequent real need for screening procedures before undergoing biological therapies.

\section{Material and methods} \label{sec:model}

In this section we formalize the scheme introduced in \cite{Wallis_math} and aimed to reproduce data of TB onset in patients treated with TNF inhibitors, with
particular attention on  infliximab (an anti-TNF mAb) and etanercept (a soluble TNF receptor). Seeking for clarity, in this section we mention only applications to the TB case, although, as we will see in the second part of the paper, the approach is rather robust and can be successfully applied to the NTM case, too.

The model, whose structure is depicted in Figure \ref{fig:markov_chain}, consists in identifying a set of possible states
for the patient subjected to biological treatments, and in fixing the likelihood for the patient to change his/her state within a proper unit time \footnote{Clearly, on large samples, some patients may experience sudden incidents
(e.g. death for other causes) or some others may assume both the drugs: the analysis has been previously purified from these cases \cite{Wallis_math,Wallis_data}.}.

The clinical states available to a test-patient are taken as follows (see Figure $1$):
\begin{itemize}
	\item[0]: Absence of infection;
	\item[1]: New infection (that after a time $\tau$ can give rise either to active TB or latent infection);
	\item[2]: Latent infection;
	\item[3]: Reactivated TB after latency;
	\item[4]: Active TB (that progress from new infection within a time $\tau$, without an intervening period of latency).
\end{itemize}
Moreover, each patient is assumed to change his/her state, following the corresponding transition probabilities, which constitute
the model parameter set, and are meant over a proper unit time $\tau$. Using $t$ to label the time, these probabilities are:
\begin{itemize}
	\item[$L$]: Probability of having a latent infection at the beginning of the observation ($t=0$),
	 while, obviously, $(1-L)$ is the probability of not having any infection at that moment;
	\item[$N$]: Probability of TB infection during the observational time;
	\item[$P$]: Probability of a new TB infection to become active TB after a time $\tau$; as a consequence, $(1-P)$ is the probability of this new infection to give rise to a latent infection;
	\item[$R$]: Probability of reactivation of a latent TB infection.	
\end{itemize}
%
We stress that such probabilistic framework is based on purely clinical variables.

On the experimental side, the available data consist in a collection of times (one for each patient) corresponding to
the onset of TB (in its active phase, namely a detectable scenario), after the beginning of the treatment with TNF blockers. As a consequence, the only states which are possible to observe are the states $3$ and $4$.
Unfortunately, as discussed in the introduction, these states (that account for ill patients) are not distinguishable one respect to the other by simply looking at the data (hence motivating both earlier studies \cite{Wallis_math,Wallis_data} and our
machinery), however, some progress can be made using stochastic extremization. The idea resembles the standard maximum likelihood and consists in finding the best values for free parameters such that the theoretical curves collapse over the experimental data \cite{Pearson}.

As already outlined, following this procedure, the main result in \cite{Wallis_math} is that the principal difference between infliximab and etanercept treatments resides in different management of latent TB, i.e. on $R$. The former seems to enhance reactivation one order of magnitude more than the latter.

\subsection*{Markov chains and master equations}

The model described in the previous section can be translated into a set of differential equations
coding for the temporal evolution  of the probability of patient's states
(which can be compared to the corresponding fractions over a sample of patients given the large
collection of data).

Being the states discrete, this can be accomplished in complete generality using Markov chains, namely a (discrete-time) probabilistic framework where the probability of being in a given state at a given time $t$ depends only on the probability distribution over all the states at the previous time step $t-1$, and on the transition rates linking these states.

\begin{figure}[!h]
\begin{center}
\includegraphics[width=4in]{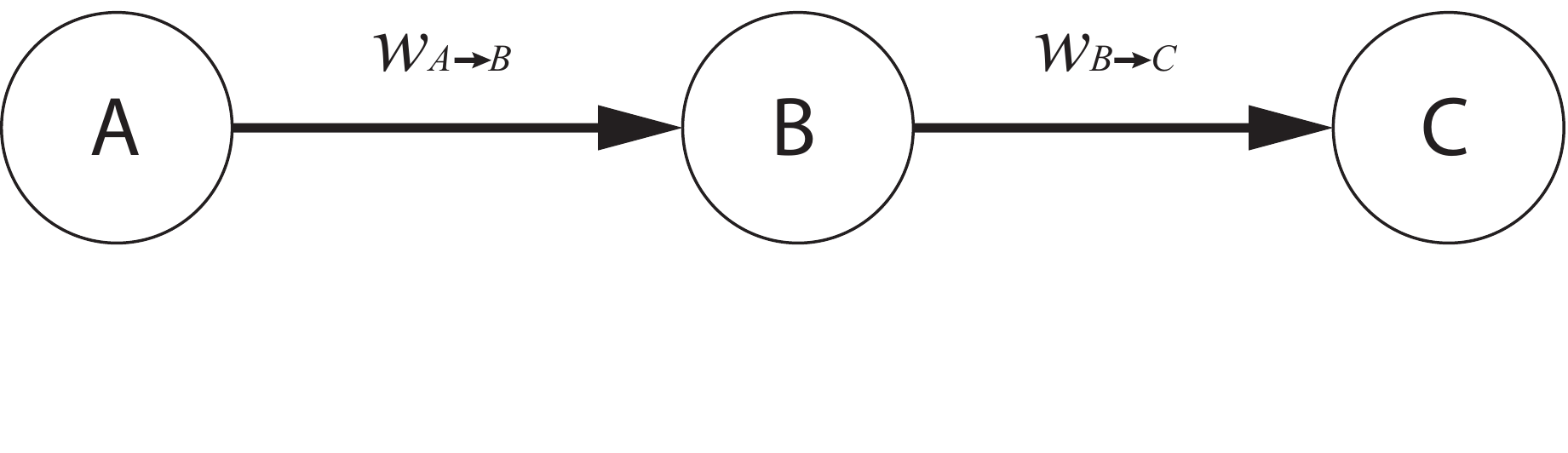}
\end{center}
\caption{
{\bf Toy Markov chain.}  From the state $A$ there is a potential flux of probability at rate $w_{A \rightarrow B}$ toward
the state $B$, hence we expect that, after a proper amount of time, a fraction of the probability $p$ will be drained from $A$
to $B$. The same holds for the situation linking $C$ to $B$. After an infinite time the probability of having the patient
in the state $C$ is one, while it is zero for the states $A, B$.}
\label{fig:markov_toy}
\end{figure}

It is instructive to consider the illustrative Markov chain with only three states ($A$, $B$ and $C$),
non-null transition rates $w_{A \to B}$ and $w_{B \to C}$
and time step $\Delta t$, shown in Figure \ref{fig:markov_toy}.

Note that, in the model, the probabilities of going from $A$ to $B$ and from $B$ to $C$ exist but not the opposite ($w_{B \to A}=0,\ w_{C \to B}=0$) hence, if the initial state is all concentrated in $C$, there will be no evolution, while if the starting point is spread among $A$ and $B$, after enough time, the probability distribution will be peaked on $C$ only (but in its {\it finite}  temporal evolution resides our interest).


Now, the probability of remaining-at/moving-into
the state $B$ in the time interval $\Delta t$ is given by the probability of already being in $B$
(hence $p_B(t)$) plus the probability of arriving in $B$ from $A$ times the probability of being in $A$ at the previous step (hence  $w_{A \to B}p_A(t)$) minus the probability of leaving $B$ to $C$ times the probability of being in $B$ at the previous step (hence  $w_{B \to C}p_B(t)$); this concept  can be written as follows:
\beq
p_B(t+\Delta t) = p_B(t) + p_A(t) \, w_{A \to B}\Delta t - p_B(t) \,w_{B \to C}\Delta t
\:.
\eeq
Since the mathematics for continuous variable differential equations is much more handily and does not change significantly the perspective if the time step is small with respect to the global time window\footnote{Both the experimental data set considered here (for TB cases \cite{Wallis_data} and for NTM cases \cite{Winthrop}) fulfill this requirement.}, we are allowed to consider the time as a continuous variable. This can be achieved straightforwardly starting from the previous equation using a limit procedure:
$$
\lim_{\Delta t \to 0} \frac{p_B(t+\Delta t) - p_B(t)}{\Delta t} = \frac{dp_B(t)}{dt} = p_A(t) w_{A \to B}  - p_B(t)w_{B \to C}.
$$
The evolution for the probability $p_B(t)$  is then ruled by the following differential equation, namely a ``Master equation", which acts as a continuous counterpart of the Markov chain in the discrete-time case:
\beq
\frac{d p_B(t)}{d t} = p_A(t) w_{A \to B}  - p_B(t)w_{B \to C}
\:.
\eeq
In general, for a system which can be in one of $M$ arbitrary states, we need a $M\times M$ transition-rate matrix $w$
(where $w_{i \to j}$ is the rate for the transition from state $i$ to state $j$) and the Master
 equation takes the form
\beq
\dot{p}_i(t) \equiv \frac{d p_i}{d t} = \sum_{j=1}^M w_{j \to i} \, p_j(t)  - \sum_{j=1}^M w_{i \to j} \, p_i(t)
\:.
\eeq
Note that we use the symbol $\cdot$ above the functions meaning their temporal derivative.
Finally, we switch to a form where the explicit timescale $\tau$ of the process appears directly in the equation, that is
\begin{equation}
\tau \dot{p}_i(t) = \sum_{j=1}^M W_{j \to i} \, p_j(t)  - \sum_{j=1}^M W_{i \to j} \, p_i(t),
\end{equation}
where $W_{i \to j} \sim w_{i \to j} \tau$ stands for the probability of transition from state $i$ to $j$ along the time interval $\tau$.

\subsection*{Master equations for the model}
\begin{figure}[!h]
\begin{center}
\includegraphics[width=4in]{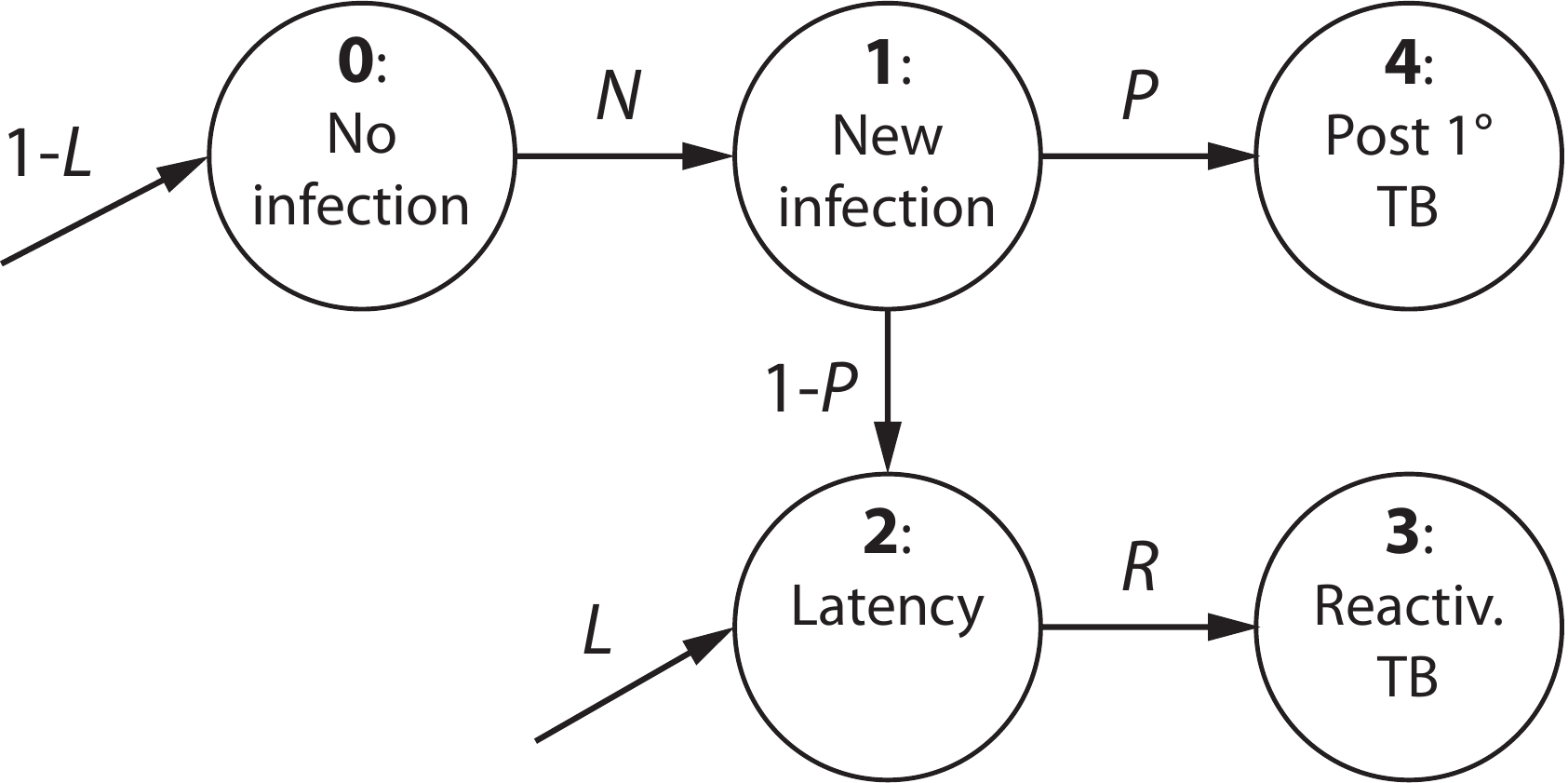}
\end{center}
\caption{
{\bf Symbolic representation of the Markov chain under investigation.} States are represented as circles and numbered from $0$ to $4$ according to the scheme outlined in the Material and Methods section. The arrows $N,P,1-P,R$ represent the transition probabilities connecting two different states, while $1-L$ and  $L$ represent the initial conditions on the states $0,2$.}
\label{fig:markov_chain}
\end{figure}

Keeping in mind Figure \ref{fig:markov_chain}, we can write down the system of differential equations describing the evolution of the five states earlier introduced as follows:
\begin{equation} \label{eq:system}
\begin{cases}
\tau \dot{p}_0 = -N \, p_0(t), \\
\tau \dot{p}_1 = N p_0(t) - p_1(t),\\
\tau \dot{p}_2 = (1-P) \, p_1(t) - R \, p_2(t),\\
\tau \dot{p}_3 = R \, p_2(t),\\
\tau \dot{p}_4 = P \, p_1(t),\\
\end{cases}
\end{equation}
with initial conditions
\begin{equation}
\begin{cases}
p_0(t=0) = 1-L,\\
p_2(t=0) = L,\\
p_1(t=0) = p_3(t=0) \, = \, p_4(t=0)\, = \, 0.
\end{cases}
\end{equation}
The numbers indexing the probabilities mirror the enumeration of the previous section, that is, $p_0$ stands for
the probability that a patient has never been affected by the infection, and so on. The parameter $\tau$ represents the typical time for a patient experiencing a new infection to either develop the disease or to fall into a latent state and it should be chosen according to the natural time-scale of the process described.
For instance, for the TB case, the data collected suggest that $\tau$ is order of a few months \cite{Wallis_data,Iseman-2008AJRCC}, and we set $\tau= 1\, month$ for the sake of simplicity and in agreement with previous works \cite{Wallis_math,Wallis_data}.

Note that, as patients affected by active TB do not start RA therapy, we set $p_1(t=0) = p_3(t=0) =  p_4(t=0) =  0$. Furthermore, the parameter $L$ tunes the initial amount of latent-TB patients with respect to free-TB patients, such that for $L=0$ all patients are healthy, while for $L=1$ all patients display a latent TB infection; as we have no ways to discriminate between healthy and latent-infected patients, $L$ is taken as a free parameter which can be estimated a posteriori comparing the solution of (\ref{eq:system}) with available data.

The solution of the system (\ref{eq:system})  can be easily obtained using first order ordinary differential equations theory
and reads off as
\begin{equation}\label{soluzione}
\begin{cases}
p_0(t) = (1-L) \, e^{-Nt/\tau},\\
p_1(t) = N \frac{1-L}{1-N} \left( e^{-Nt/\tau} -e^{-t/\tau} \right),\\
p_2(t) = - \frac{(1-P)(1-L)N}{(1-N)(N-R)} \, e^{-Nt/\tau} + \left[ L + \frac{(1-P)(1-L)N}{(N-R)(1-R)} \right] \, e^{-Rt/\tau} +
		\frac{(1-P)(1-L)N}{(1-N)(1-R)} \, e^{-t/\tau} 		,\\
p_3(t) = L+(1-P)(1-L) + \frac{R(1-P)(1-L)}{(1-N)(N-R)}  e^{-Nt/\tau} -
		\left[ L + \frac{(1-P)(1-L)N}{(1-R)(N-R)} \right] e^{-Rt/\tau}\\
		\;\;\;\;\;\;\;\;\;\; - \frac{(1-P)(1-L)RN}{(1-N)(1-R)}
		e^{-t/\tau} ,\\
p_4(t) = P \frac{1-L}{1-N} \left(1- N  -  e^{-Nt/\tau} + N\,e^{-t/\tau} \right).
\end{cases}
\end{equation}
Of course, since the total amount of patients is conserved, $C_0=p_0+p_1+p_2+p_3+p_4$ is an integral of motion, that is
\begin{equation}
0 = \frac{d}{dt} C_0  \Rightarrow C_0 = p_0(t) +p_1(t)+ p_2(t) + p_3(t) +p_4(t) = \mathrm{const}.
\end{equation}
Beyond $C_0$, the system ($\ref{eq:system}$) admits another integral of motion $C_1$, namely
\begin{equation}
0 = \frac{d}{dt} \left[ p_2(t)+p_3(t)+\frac{P-1}{P} p_4(t) \right] \Rightarrow C_1 = p_2(t)+p_3(t)+\frac{P-1}{P} p_4(t)= \mathrm{const}.
\end{equation}
This means that the rate of growth for patients in the latency branch (i.e. in states $2, 3$) equals the rate of growth for the rest of infected patients (i.e. in state $4$) weighted by a factor $P^{-1} -1$, so that the smaller $P$ and the larger the difference between the related rates.
The knowledge of integrals of motion can be very useful as they allow to obtain information in a very simple way; for instance, should $P$ drop, then $p_4(t)$ would also decrease (or, analogously, $p_2(t) + p_3(t)$ would increase) in order to maintain $C_1$ constant. Given $C_0$ and $C_1$, other integrals of motion, which are combination of $C_0$ and $C_1$, can be trivially built. For example, $C_2 = C_0 - C_1$ fulfills
\begin{equation}
0 = \frac{d}{dt}\left [ p_1(t)+p_0(t)+\frac1P p_4(t) \right ] \Rightarrow C_2 = p_1(t)+p_0(t)+\frac1P p_4(t)= \mathrm{const}.
\end{equation}
We underline that this kind of investigation can be accomplished only through an analytical study of the system.

As discussed above, the fraction of active TB cases is given by the sum of the fraction of cases of direct TB after infection and of the fraction of cases with reactivated TB; namely, calling $c(t)$ the total fraction of cases, we have:
\beq \label{cc}
c(t) = p_3(t)+p_4(t)
\:.
\eeq
In the above quantity, the time dependence appears only through three different exponential decay terms $(e^{-Nt/\tau}, \ e^{-Rt/\tau} ,\ e^{-t/\tau})$, which vanish at infinite time, so that the solution becomes a constant term equal to 1, meaning that, if we wait for a sufficient long (possibly infinite) time, all patients become sick (although, obviously, they can possibly die earlier due to reasons not related to RA/TB). In order to deepen the temporal evolution of these probabilities at relatively short times, it is useful to use a little bit of algebraic manipulation to distinguish constant terms from decaying terms, in such a way that we get
\beq\label{eq:c}
c(t) = 1 + k_1 e^{-t/\tau}  + k_R e^{-Rt/\tau} + k_N e^{-Nt/\tau},
\eeq
where the three constants $k_1$, $k_R$ and $k_N$ are related to the physiological parameters by
\begin{equation}
\begin{cases}
k_1 = \frac{N(1-L)(P-R)}{(1-N)(1-R)},\\
k_R = -L +\,\frac{N(1-P)(1-L)}{(1-R)(R-N)},\\
k_N = -\frac{(1-L)(R-PN)}{(1-N)(R-N)}.
\end{cases}
\end{equation}
Of course, from $c(t)$ one can derive the effective number of cases multiplying $c(t)$ by the overall number of treated patients.

Before turning attention to the fitting procedure, we stress that the analytical solution in Eq.~(\ref{soluzione}) was successfully checked through numerical methods, i.e. fourth-order Runge-Kutta algorithm and Monte Carlo simulations\footnote{Here, with Monte Carlo simulation we mean a simulation in which a set of virtual patients evolves in time following the Markov chain of Figure \ref{fig:markov_chain} giving a sample of the evolution of the fraction of cases during time. In our simulations we set $10^6$ virtual patients.
}.


\section{Results}\label{sec:approx}

\subsection*{The TB-infection case}
Having obtained the complete solution of the model and exploiting the available information on parameters,
we now look for proper approximations able to highlight the effective behavior of $c(t)$ in cases of practical interest, starting
with the TB scenario.

In particular, Eq.~(\ref{eq:c}) can be reduced to a simpler form if we note that the probability $N$ of TB infection is much smaller than all the other parameters, in agreement with studies on TB and with results found in \cite{Wallis_math} and, a posteriori, in the current paper (see Tab.~1).

Hence, as a first approximation step, we assume $N \ll 1$ and $N/R \ll 1$ such that we can expand the solution,
at the first order in $N$ and $N/R$, as follows\footnote{Here and in the following we use the ``big-O'' Landau  notation to characterize the growth rate of functions; more precisely, being $f(x)$ and $g(x)$ two arbitrary functions, we say $f(x) = \mathcal{O}(g(x))$ as $x \to 0$ if there exists a positive real number $M$ such that $|f(x)| \leq M |g(x)|$.}:
\bea
k_1 &=&  (1-L) \left(P-R \; \;  \frac{ 1-P }{ 1-R } \right) N [1 + N + \mathcal{O}(N^2)], \nonumber\\
k_R &=& -L +\, \frac{(1-P)(1-L)}{R(1-R)} \; N \left[ 1+ \frac{N}{R} + \mathcal{O}\left ( \frac{N^2}{R^2} \right)  \right]
\:,\nonumber\\
\nonumber
k_N &=& -(1-L) \left \{ P [1 + N + \mathcal {O}(N^2)] + (1-P) [1 + N + \mathcal {O}(N^2)] \left [1+ \frac{N}{R} + \mathcal{O}\left ( \frac{N^2}{R^2} \right) \right ] \right \}.
\eea
Therefore, with some algebra and retaining only up-to-linear terms in $N$ or $N/R$, we get
\bea\label{eq:c0c1}
\nonumber
c(t) &\approx&  1 + N (1-L)  \frac{P-R}{1-R}   e^{-t/\tau} + \left[ - L + \frac{(1-P)(1-L)}{1-R} \frac{N}{R} \right] e^{-Rt/\tau} \\
&-& (1-L) \left [1+N +\frac{N}{R} (1-P)\right] e^{-Nt/\tau}.
\eea

\begin{figure}[!h]
\begin{center}
\includegraphics[width=4in]{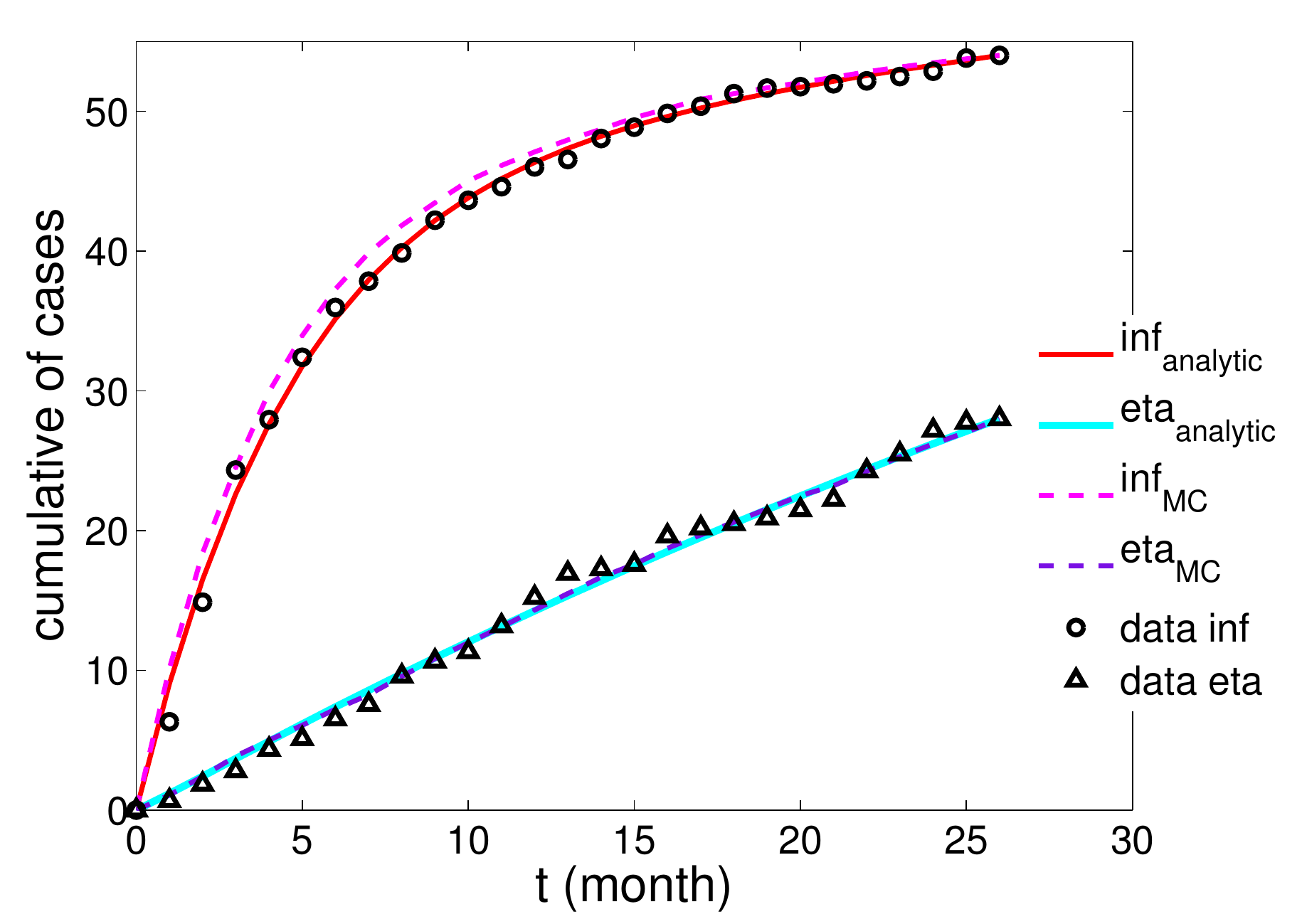}
\end{center}
\caption{
{\bf Comparison between experimental, analytical and numerical results for the TB-infection case.}  Cumulative number of patients undergoing active TB-infection. Experimental data from \cite{Wallis_math} ($\bullet$ for infliximab-treated patients and $\triangle$ for etanercept-treated patients) are compared with the approximated analytical solution (see Eq.~\ref{eq:prelinear} and Eq.~\ref{eq:linear}, respectively, solid curves) and with data from numerical simulation (dashed curves). The parameters used to draw the analytical curves correspond to the best-fit coefficients and are reported in Tab.~1. 
Notice that here we consider the extensive number of patients affected by TB over a population of $\mathcal{N} = 10^5$ treated patients, according to experimental results.}
\label{cumulative_TNF}
\end{figure}

\begin{figure}[!h]
\begin{center}
\includegraphics[width=4in]{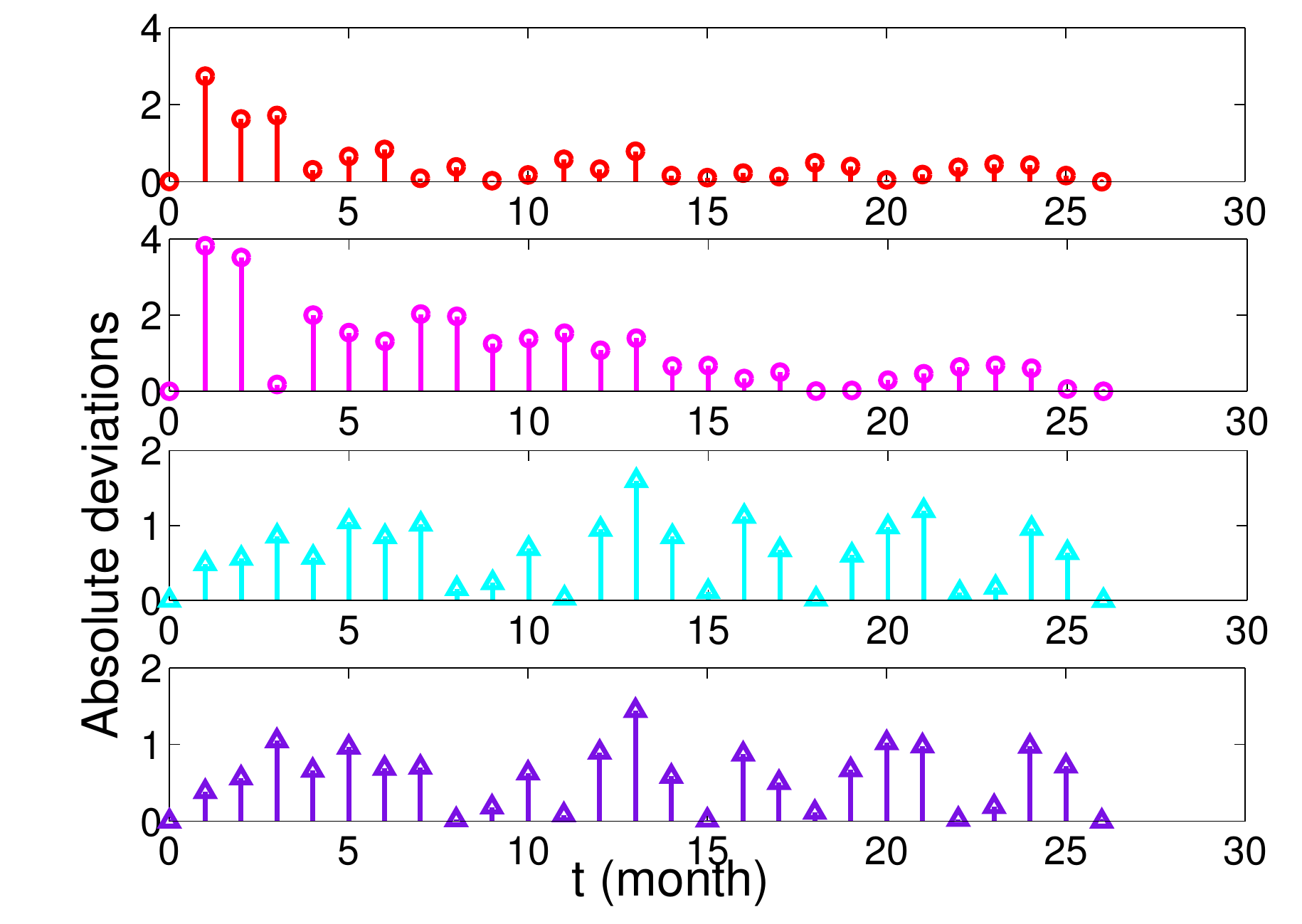}
\end{center}
\caption{
{\bf Deviations between experimental, analytical and numerical results for the TB-infection case.}  Absolute difference between experimental data and theoretical data reported in Fig.~3; the same legend holds.
}
\label{ritardati}
\end{figure}

Let us now move further and focus on the exponential terms. First, we notice 
that 
$1 > R > N$ and, consequently, we can neglect the term $e^{-t/\tau}$, as it decays much faster that both $e^{-Rt/\tau}$ and $e^{-Nt/\tau}$. Moreover, since the time range considered is $\approx 30$ months and $N$ is expected to be $\ll \tau/t \approx 10^{-2}$,
we can expand $e^{-Nt/\tau}$ as $e^{-Nt/\tau} \approx 1 - Nt/\tau$, and considering only the leading dependence on $t$, we get
\begin{equation}\label{eq:prelinear}
c(t) \approx 1 - (1-L) \left [1+N +\frac{N}{R} (1-P)\right] +  \left[ - L + \frac{(1-P)(1-L)}{1-R} \frac{N}{R} \right] e^{-Rt/\tau}  .
\end{equation}
As for $e^{-Rt/\tau}$, a similar approximation ($e^{-Rt/\tau} \approx 1-Rt/\tau$) can be adopted as long as $R \lesssim
 10^{-2}$, so to obtain the following linear approximation
\beq\label{eq:linear}
c(t) \approx 1+ k_R (1-Rt/\tau) + k_N.
\eeq
Notice that a smaller (larger) estimate for $\tau$ would simply require a stricter (softer) condition on $N$ and on $R$ for the related linear expansions to hold (on the same time range); the model would not be affected and the parameters coupled with time, i.e. $N,R$, would be accordingly rescaled.
As shown in Figure \ref{cumulative_TNF}, the approximation (\ref{eq:linear}) is rather good only for
etanercept-treated patients, for which the best fit yields $R = 2.24 \cdot 10^{-2}$. On the other hand,
if we consider infliximab-treated patients, the approximation (\ref{eq:linear}) does not fit data, while using (\ref{eq:prelinear}) we get a good overlap with data and the best fit yields $R=2.12 \cdot 10^{-1}$, confirming that now $Rt/\tau$ is no longer small over the time window. All best-fit coefficients are reported in Tab.~$1$, and the related errors are shown in Figure \ref{ritardati}.

We can estimate how sensitive $c(t)$ is with respect to the system parameters by deriving its analytic expression (see Eq.~\ref{eq:c}) with respect to $N, P, R, L$, respectively; in this way we get that, in the regime  $N \ll (1, P, R, L)$, the most relevant parameter affecting the behavior of $c(t)$ is $R$.
Another argument in favor of this claim is that, in the zero approximation of the solution (i.e. neglecting even terms $\mathcal{O}(N)$), $P$ does not appear at all.

\begin{figure}[!h]
\begin{center}
\includegraphics[width=4in]{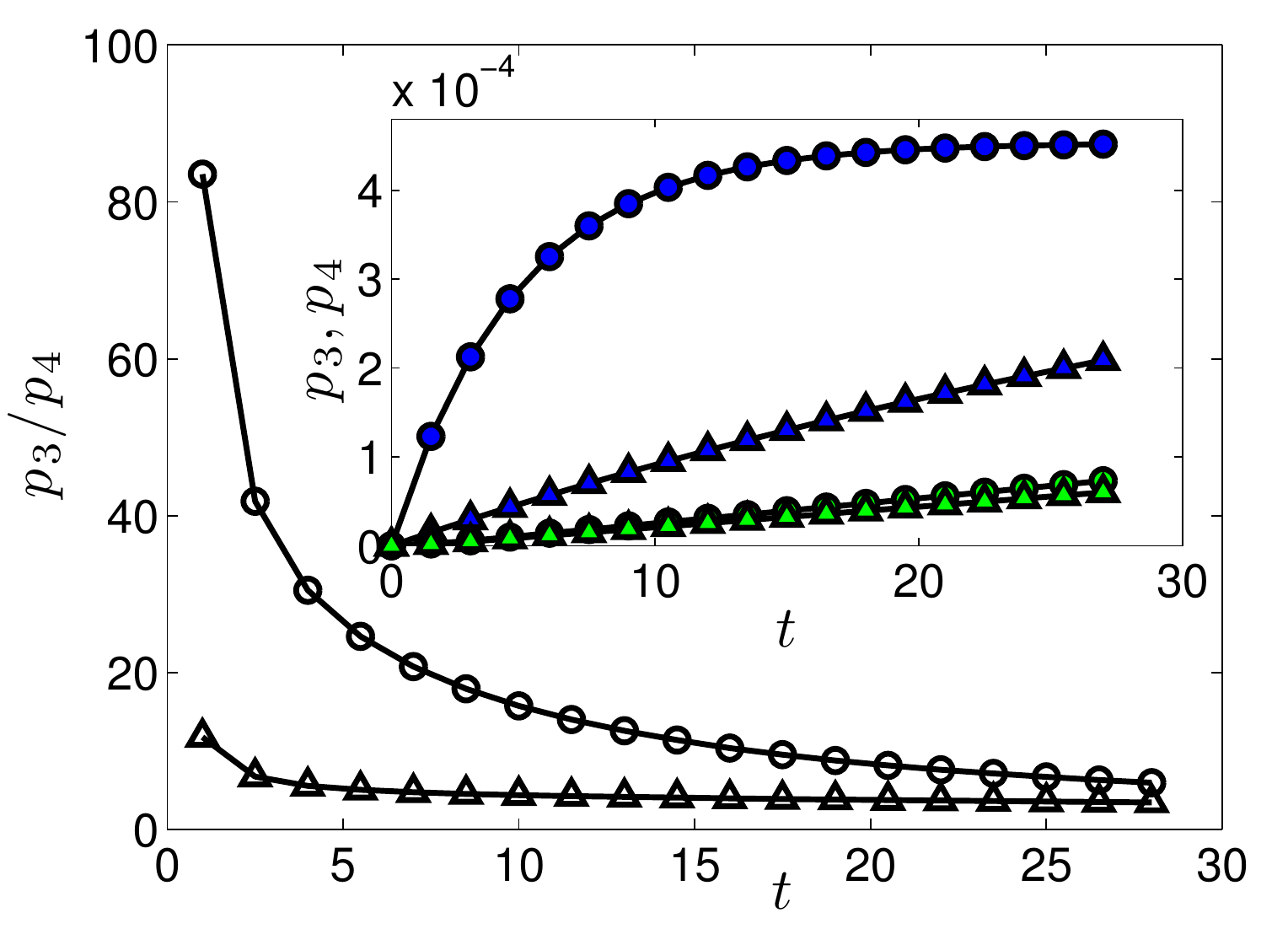}
\end{center}
\caption{
{\bf Probabilities $p_3$ and $p_4$ versus time for the TB-infection case.}  The probability $p_3$ of being in state $3$ (TB reactivation) and $p_4$ of being in state $4$ (Post $1^{\circ}$ TB) are calculated from Eq.~\ref{soluzione}, using the parameters of Tab.~$1$. In the main plot we compare the ratio $p_3/p_4$ obtained from infliximab ($\bullet$) and etanercept ($\triangle$) parameters. In the inset, we depict each single probability, namely $p_3$ for infliximab (dark $\bullet$) and for etanercept (dark $\triangle$), $p_4$ for infliximab (bright $\bullet$) and for etanercept (bright $\triangle$); notice that the two sets of data for $p_4$ are partially overlapped. Lines are guide for the eye.
}
\label{fig:refer}
\end{figure}

In order to get further insight on the effect of infliximab and of etanercept on TB incidence, in Fig.~\ref{fig:refer} we plotted the model predictions for the percentage of patients having TB because of new infections ($p_4(t)$) or reactivations ($p_3(t)$).

To summarize, our results confirm that, in the present context, the most important difference between therapies based on infliximab or etanercept is that the former enhances TB reactivation more than the latter, in fact, we found $R_{INF} \sim 10 \, R_{ETA}$, in agreement with \cite{Wallis_math}.
Such a discrepancy implies even a qualitatively different behavior of $c(t)$ over the time-window considered: the number of infliximab-treated patients experiencing a TB infections grows exponentially in time, while for etanercept-treated patient the growth is linear.

\subsection*{The NTM-infection case}
The analytical expression for $c(t)$, (see Eqs.~\ref{soluzione} and \ref{cc}) holds for a general environment schematizable as in Fig.~\ref{fig:markov_chain}; in the last paragraph we used the details of TB infections to implement convenient approximations. We now turn to NTM-infections in patients treated with infliximab and etanercept and look for proper approximations able to highlight the characteristic features of such case. The experimental data we refer to are those
reported in \cite{Winthrop} and collected over the period $1999-2006$.
Overall, there were $239$ reports of NTM infections in patients
who were receiving anti-TNF therapy.
Most reports were for patients receiving infliximab ($75\%$) or
etanercept ($17\%$) and here we shall focus just on these drugs.

Some remarks are in order here to merge mathematically the TB and NTM scenarios.

First, it is important to notice that patients affected by NTM lung infections typically suffer
through long periods of illness before a clinical diagnosis is made.
To  comply with the actual state of the art on the involved time-scales we follow M. Iseman and T. Marras \cite{Iseman-2008AJRCC}
that we quote: ``Preliminary prevalence estimates have been made, assuming that the disease duration for TB
is $8$ months and for pulmonary NTM is within the range of $4$ to $10$ years'' ($48$ to $120$ months),
hence for NTM the timescale is at least one order of magnitude larger than in the TB counterpart.

Another important point is that available data on NTM-diseases related to drug therapies lack the size of the survey, namely the number of patients participating to the screening, consequently, we can quantify the (cumulative) amount of sick patients,
but we do not know their percentage.
This is not a serious deficiency since, while we do not have access to the very reactivation probabilities $R_{INF}$ or $R_{ETA}$, their ratio (which cancels out both  time-scales and survey-size) still retains a quantitative information content.

As a result of the first remark, the (average) exit time from state $1$, i.e. $\tau$, is comparable to the experimental time window, and one can expand the solution reported Eq.~\ref{soluzione} at second order in $t/ \tau$ as
\begin{eqnarray}
p_3(t) &=&
RL \frac{t}{\tau} - LR^2 \frac{t^2}{2\tau^2} + \mathcal{O}\left(\frac{t^3}{\tau^3}\right),\\
p_4(t)  &\approx& PN(1-L) \frac{t^2}{2 \tau^2} + \mathcal{O}\left(\frac{t^3}{\tau^3}\right).
\end{eqnarray}
We notice that, at this stage of expansion, $p_4(t)$ grows quadratically with time, while $p_3(t)$ presents two contributes growing, respectively, linearly and quadratically with time, the former being the leading one.

\begin{figure}[!h]
\begin{center}
\includegraphics[width=4in]{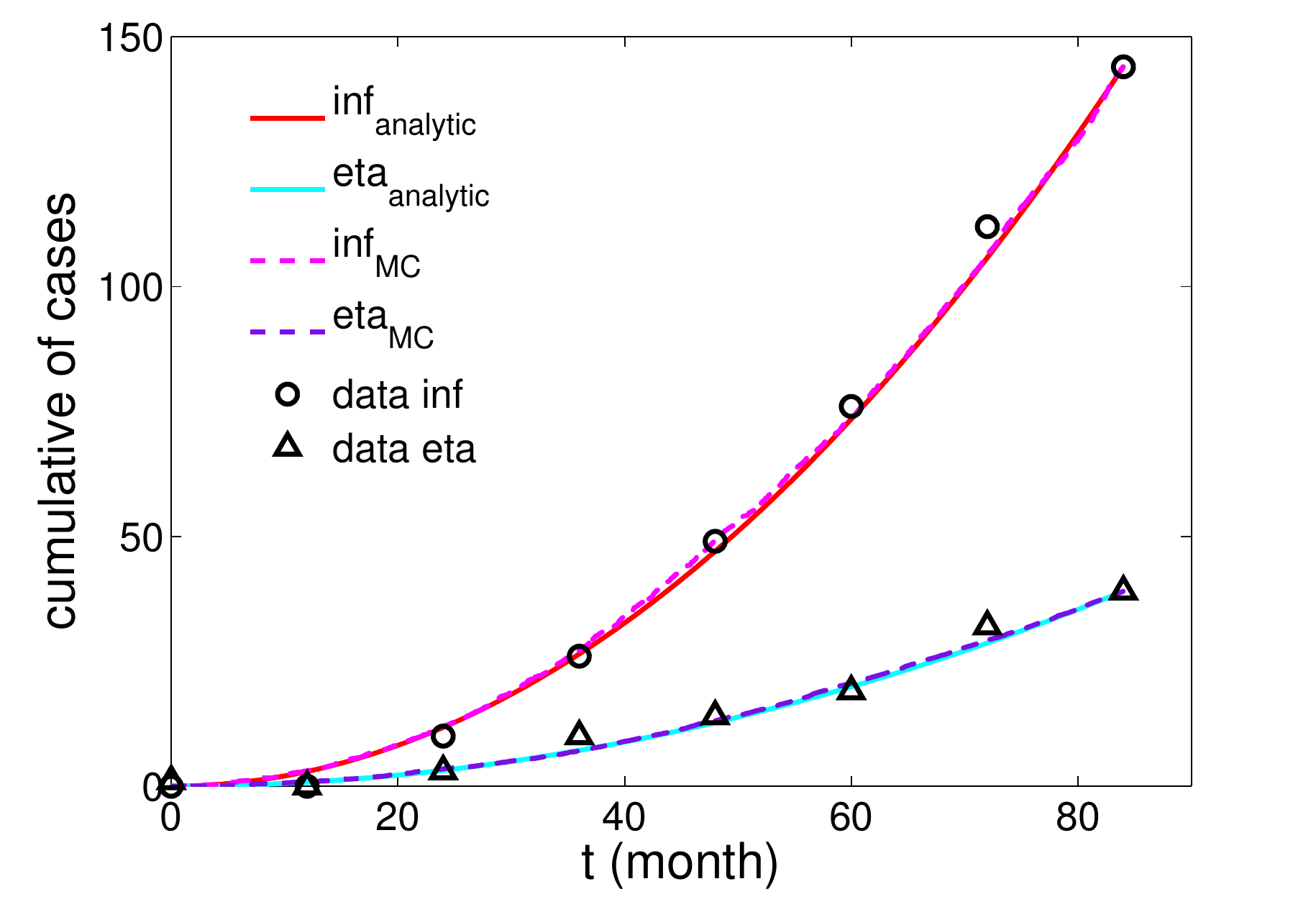}
\end{center}
\caption{
{\bf Comparison between experimental, analytical and numerical results for the NTM-infection case.}  Cumulative number of patients undergoing active NTM-infection. Experimental data from \cite{Winthrop} ($\bullet$ for infliximab-treated patients and $\triangle$ for etanercept-treated patients) are compared with the approximated analytical solution (see Eq.~$19$, solid curves) and with data from numerical simulation (dashed curves). The parameters used to draw the analytical curves correspond to the best-fit coefficient and are reported in Tab.~$2$.
Notice that here we consider the extensive number of patients affected by NTM over the whole population of treated patients, according to experimental results.
}
\label{cumulative_NTM}
\end{figure}

\begin{figure}[!h]
\begin{center}
\includegraphics[width=4in]{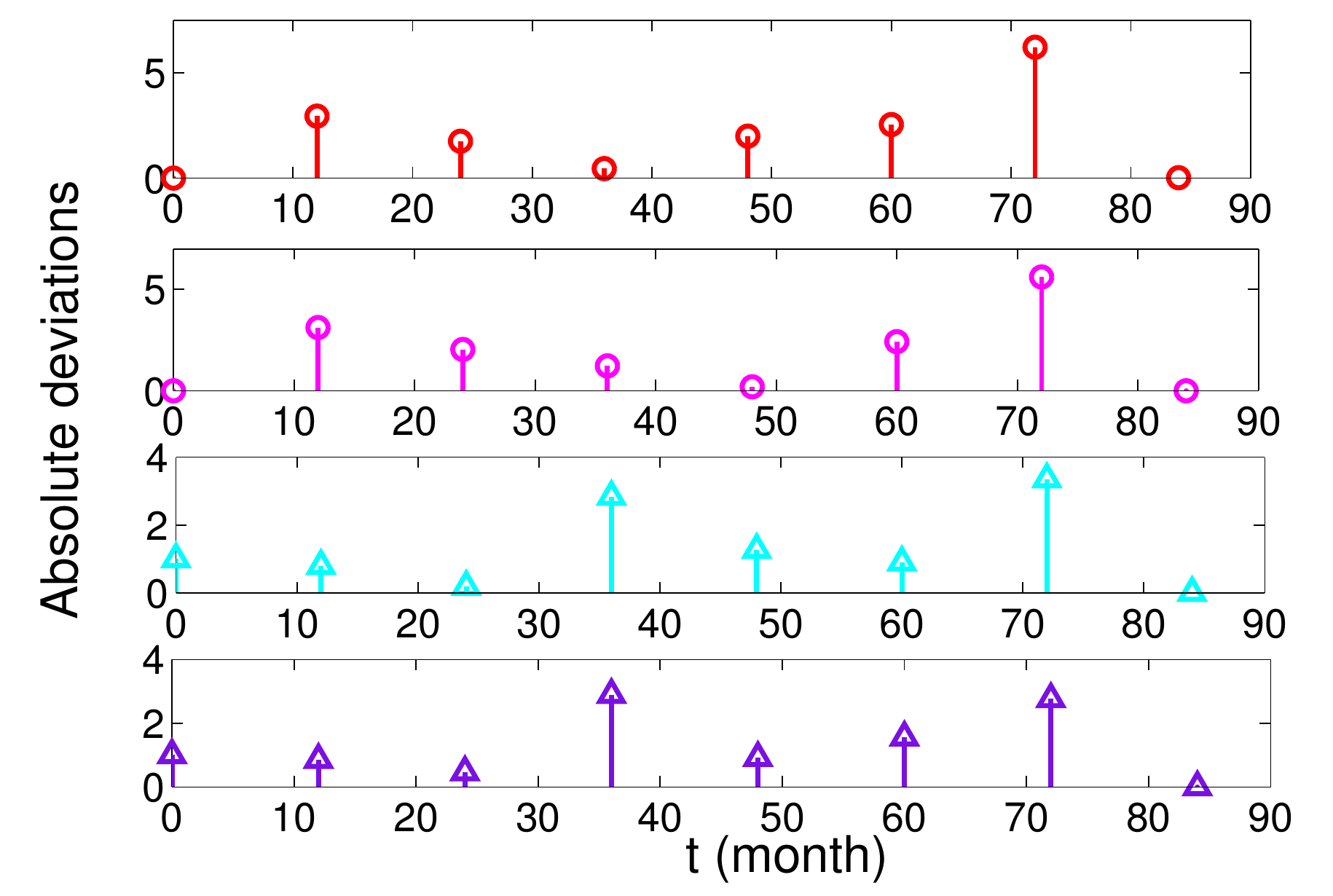}
\end{center}
\caption{
{\bf Deviations between experimental, analytical and numerical results for the NTM-infection case.}  Absolute difference between experimental data and theoretical data reported in Fig.~5; the same legend holds.
}
\label{scarti_NTM}
\end{figure}

Such expressions must now be compared with experimental data, which display a purely quadratic growth in time,
i.e. compatible with $c(t) = a t^2$ (see Fig.~\ref{cumulative_NTM}).
In order for the comparison to hold, one must therefore drop the linear coefficient, and this implies $R \sim 0$ (or $L \sim 0$ which is conceptually pretty similar). With this choice of parameters we get
%
%
\begin{equation}\label{sol_NTM}
\begin{array}{c}
p_3(t) \approx 0 \\
p_4(t)  \approx PN \frac{t^2}{2 \tau^2}  \\
\end{array} \Rightarrow c(t) \approx PN \frac{t^2}{2 \tau^2},
\end{equation}
and this form successfully fits experimental data, as shown in Figure \ref{cumulative_NTM}; the best-fit coefficients are reported in Tab.~2, and the related errors are shown in Figure \ref{scarti_NTM}.

The functional form of $p_4(t)$ highlights why finding an estimate of the parameters
is extremely difficult  through blind numerical extremization, without any explicit analytical hint:
The latency-branch turns out to be negligible in NTM-context and this
results in a coupling of the parameters $P$ and $N$ as only a single probability streaming toward $p_4(t)$ survives. Therefore the two parameters alone
are undetectable.

Using the coefficients of Tab.~2 in numerical simulations we obtain results in excellent agreement with real data and analytical ones.
Notice that, impressively, the fit is very good although here the optimization relies only on one parameter.

We stress that the qualitative difference between the behaviors of $c(t)$ in the case of TB and NTM infection (see Eqs.~\ref{eq:prelinear} and \ref{sol_NTM}) mainly stems from a significant gap in the related disease duration (few months versus several years). In the latter case $\tau$ is relatively large to allow the expansion of the exponential functions into a polynomial form, while for TB this can be accomplished only under the condition of small enough $R$. Such expansions do not significantly alter the predicted values of the fit parameters, as long as the underlying assumptions are consistent with facts.

To summarize, the theoretical framework developed evidences that the rate of reactivation $R$ is vanishing:
This issue makes the fraction $p_3$ of reactivation cases negligible, while the number of activation cases grows
quadratically with time through the contribution of only $p_4$, consistently with the actual understanding of NTM-pathology achieved
through standard pathways.
Moreover, differently from the TB case, in the NTM scenario no qualitative
difference can be detected between infliximab and etanercept: The parabolic behavior for $c(t)$ seems robust.

\section{Discussion}
In the last decades, several tools and concepts stemmed from the
fields of stochastic mathematics and theoretical physics have been proposed to help the investigation of  the
immunological world, ranging from kinetic theories \cite{bellomo}, to associative
neural networks \cite{Agliari-JBD2012,Agliari-JTB2011}, to cellular automata and more \cite{paulsson, cooper}.

Along the same line, in these notes we formalized and extended a stochastic approach to data
analysis (originally
introduced in \cite{Wallis_bio, Wallis_lancet,Wallis_math,Wallis_data}) for evidencing underlying correlations between adverse events and therapies based on immunosuppressants.
In particular, the focus of our investigation concerns the risk
of reactivation of latent mycobacterial infection in patients undergoing treatment with TNF-inhibitors.

We gave a clear and complete mathematical backbone to  this approach,
building it on explicit Markov
processes, whose continuous-time limit yields the Master equation governing the evolution of the expected
fraction of patients $c(t)$ exhibiting an active infection. We also solved the Master equation in all
details finding an analytical expression for $c(t)$. Such mathematical developments make the
original approach much more versatile and general:
For instance, handling the complete (mathematical) solution allows to  better account for reasonable approximations, tackling their control quantitatively (e.g. finding the proper timescales involved in the process or the integrals of motion constraining the evolution of the system).
Furthermore,  we can finally consider, within the same
framework, different problems. In particular, we focused on TB and NTM infections emerging during anti-TNF
therapies (infliximab and etanercept) according to data reported in \cite{Wallis_math,Winthrop}.

In the former case, we  recovered previous findings showing that the rate of reactivation $R$ of
TB from a latent state to an active state plays a crucial role: being $R_{INF} \sim 10 \, R_{ETA}$ we get
qualitatively different behaviors for $c(t)$. More precisely, once fixed the observational time-window,
for infliximab $c(t)$ grows exponentially  with time, while for etanercept it grows linearly with time.
Hence, these results sustain the need, for patients candidate to TNF blockers, to perform an accurate TB screening at baseline, irrespective of the type of antiTNF. Indeed, screening may decrease the risk of TB reactivation in such patients, while it is less clear what should be done to prevent NTM disease occurrence or progression in patients taking biologic agents. Importantly, for this purpose, we found that the comparison with experimental data allows to infer that reactivation plays a very minor role for both the therapies and that $c(t)$ grows quadratically with time.

We checked our results also against Monte Carlo simulation with excellent agreement.

Furthermore, our results are all consistent with recent experimental data and seem to indicate that TB and NTM infections are sustained by different pathogenetic mechanisms.

Non-tuberculous Mycobacteria are present in large numbers in the environment, including fresh water, aerosols, biofilms, and soils \cite{Falkinham}. There are thus many opportunities for acquisition of NTM infection during ordinary daily activities, although the true incidence is not known. In contrast, nearly all transmission of Mycobacterium tuberculosis infection results from inhalation cough-generated aerosols from persons with active pulmonary TB. The annual risk of TB infection (ARTI) can be calculated from age-specific rates of tuberculin skin test reactivity; in most instances it is directly related to TB prevalence. Thus although the ARTI may reach $4\%$ in highly TB-endemic regions such as South Africa, it is as low as $0.01\%$ in much of Northern Europe and North America \cite{Sutherland,Shanaube,Daniel}. These epidemiologic findings are consistent with the results of our mathematical model, and underscore the interplay of microbial and host biology in determining the relative contributions of reinfection and reactivation to mycobacterial pathogenesis.

Hence, while the screening for TB is necessary prior to initiating biologics, when considering NTM only a watchful monitoring during the treatment is recommended.
This finding is particularly relevant, since it allows to avoid screening for NTM infection, which is complicated by the poor sensitivity of chest radiograph and more expensive and invasive techniques, such as chest computed tomography scan and/or bronchoscopy, should be used.

It is worth stressing that this methodology, being based on very standard stochastic procedures, has the advantage to
 hold beyond the test case of Mycobacteria.
We hope that this test-case may shed light to future developments of this sideline approach in figuring out
adverse events of biological therapies.



\clearpage
\section*{Figure Legends}

\clearpage
\section*{Tables}

\begin{table}[!h] 
 \begin{center}
 \begin{tabular}{l|| l|}
     & {\bf TB-infection case: Best-fit coefficients.} \\
   \hline
   \hline
  ${\bf L}$ &  $( 4.52 \pm 0.65) \cdot 10^{-4}$  \\
  \hline
  ${\bf N}$ & $(2.88 \pm 0.23) \cdot 10^{-6}$ \\
  \hline
  ${\bf R_{\mbox{\tiny INF}}}$  & $(2.12 \pm 0.19) \cdot 10^{-1}$ \\
  \hline
  ${\bf P_{\mbox{\tiny INF}}}$  & $(9.76 \pm 0.82) \cdot 10^{-1}$ \\
  \hline
  ${\bf R_{\mbox{\tiny ETA}}}$  & $(2.24 \pm 0.15) \cdot 10^{-2}$ \\
  \hline
  ${\bf P_{\mbox{\tiny ETA}}}$  & $(8.03 \pm 0.71) \cdot 10^{-1}$ \\
  \hline
 \end{tabular}
 \end{center}
\begin{flushleft} \textbf{Tab. 1. TB-infection case: Best-fit coefficients.} Best fit parameters obtained through the maximum likelihood method (see Eqs.~15, 16, respectively).  Here we used $\tau = 1 \, month$, consistently with clinical data.
The fit was accomplished with the constraint that the parameters $N$ and $L$ are the same for both therapies, as they are drug-independent. The average relative error on these parameters is $\approx 9 \%$.
\end{flushleft}
 \label{tab:fit}
\end{table}

\begin{table}[!h] 
 \begin{center}
 \begin{tabular}{l|| l|}
     & {\bf NTM-infection case: Best-fit coefficients.} \\
   \hline
   \hline
  ${\bf N P_{\mbox{\tiny INF}}}$  & $(5.88 \pm 0.46) \cdot 10^{2}/\mathcal{N}$ \\
  \hline
  ${\bf N P_{\mbox{\tiny ETA}}}$  & $(1.59 \pm 0.13) \cdot 10^{2}/\mathcal{N}$ \\
  \hline
 \end{tabular}
 \end{center}
\begin{flushleft} \textbf{Tab. 2. NTM-infection case: Best-fit coefficients.} Best fit parameters obtained through the maximum likelihood method (see Eq.~19). Here we used $\tau = 10 \, years$, consistently with clinical data. Notice that here parameters only count through the combination $PN$, and, again, $N$ was thought as drug-independent. Moreover, in this case available data are extensive, so that the explicit number of overall treated patients $\mathcal{N}$ has been introduced. The average relative error on these parameters is $\approx 8 \%$.
\end{flushleft}
 \label{tab:fit2}
\end{table}

\clearpage
\bibliographystyle{unsrt}
\bibliography{biblioTNF}

\end{document}